\documentclass[11pt,journal, comsoc]{IEEEtran}
\usepackage{times}
\usepackage{booktabs} 
\usepackage{tikz} 
\usepackage{graphicx}
\usepackage{amsmath}
\usepackage{amsthm}
\usepackage{algorithm}
\usepackage{algorithmic}
\usepackage{mathtools,amssymb}
\usepackage{multirow}
\usepackage{enumitem,kantlipsum}
\usepackage{caption}
\usepackage{subcaption}
\usepackage{cite}
\usepackage{ulem}
\usepackage{pifont}

\newcommand{\blue}[2]{{{#1}}}
\newcommand{\Repone}[2]{{#1}}
\newcommand{\Reptwo}[2]{{#1}}
\newcommand{\Repthree}[2]{{#1}}

\title{PP-MARL: Efficient Privacy-Preserving Multi-agent Reinforcement Learning for Cooperative Intelligence in Communications}

\author{Tingting Yuan,
        Hwei-Ming Chung,
        and~Xiaoming Fu
\IEEEcompsocitemizethanks{
\hspace{-1.2em}
Tingting Yuan and Xiaoming Fu are with the Computer Networks Group, Georg-August-University of Göttingen, Göttingen, Germany.
Hwei-Ming Chung is affiliated with NOOT Tech. Co., Ltd., China.\\
\IEEEcompsocthanksitem Corresponding authors: 
Tingting Yuan, tingt.yuan@hotmail.com}

}

\begin{document}
\maketitle

\begin{abstract}
Cooperative intelligence (CI) is expected to become an integral element in next-generation networks because it can aggregate the capabilities and intelligence of multiple devices.
Multi-agent reinforcement learning (MARL) is a popular approach for achieving CI in communication problems by enabling effective collaboration among agents to address sequential problems.
However, ensuring privacy protection for MARL is a challenging task because of the presence of heterogeneous agents that learn interdependently via sharing information.
Implementing privacy protection techniques such as data encryption and federated learning to MARL introduces the notable overheads (e.g., computation and bandwidth).
To overcome these challenges, we propose PP-MARL, an efficient privacy-preserving learning scheme for MARL.
PP-MARL leverages homomorphic encryption (HE) and differential privacy (DP) to protect privacy, while introducing split learning to decrease overheads via reducing the volume of shared messages, and then improve efficiency.
We apply and evaluate PP-MARL in two communication-related use cases.
Simulation results reveal that PP-MARL can achieve efficient and reliable collaboration with 1.1-6 times better privacy protection and lower overheads (e.g., 84-91\% reduction in bandwidth) than state-of-the-art approaches.
\end{abstract}

\begin{IEEEkeywords}
Cooperative Intelligence, Privacy Preservation, Multi-agent Reinforcement Learning
\end{IEEEkeywords}

\section{Introduction}\label{sec:intro}
\Repone{

Cooperative intelligence (CI) \cite{dafoe2021cooperative, dafoe2020open} is expected to facilitate next-generation networks by establishing collaboration among various communication-related intelligent equipment.
Multi-agent reinforcement learning (MARL) is a popular approach for achieving CI in addressing sequential problems in communication \cite{zhang2019multi}, such as adaptive routing and resource allocation \cite{yuan2020dynamic}. Because MARL enables effective collaboration among agents through cooperative learning, leading to more efficient and effective communication.}{1, 5, 6}{}


Privacy concern is a prominent issue for MARL because it heavily relies on information sharing.
For example, in aerial base station control, drones need to move cooperatively to achieve given goals (e.g., offer emergency communication).  
Due to privacy concerns, drones may be unwilling to share their data/information directly for cooperation (e.g., in-drone videos and movement trajectory).
Another example is collaborative network management with edge controllers \cite{yuan2020dynamic}, in which edge information (e.g., customer locations, requirements, and feedback) is private.
As a result, privacy issues become a major obstacle to the deployment of MARL to achieve CI.
\Reptwo{
However, privacy protection in MARL is complex due to interdependent learning among heterogeneous agents, which differ in observation range, action space, learning algorithms, communication, and objectives.
}{1}{}

\begin{figure*}[t!]
\centering
\includegraphics[width=0.85\textwidth] {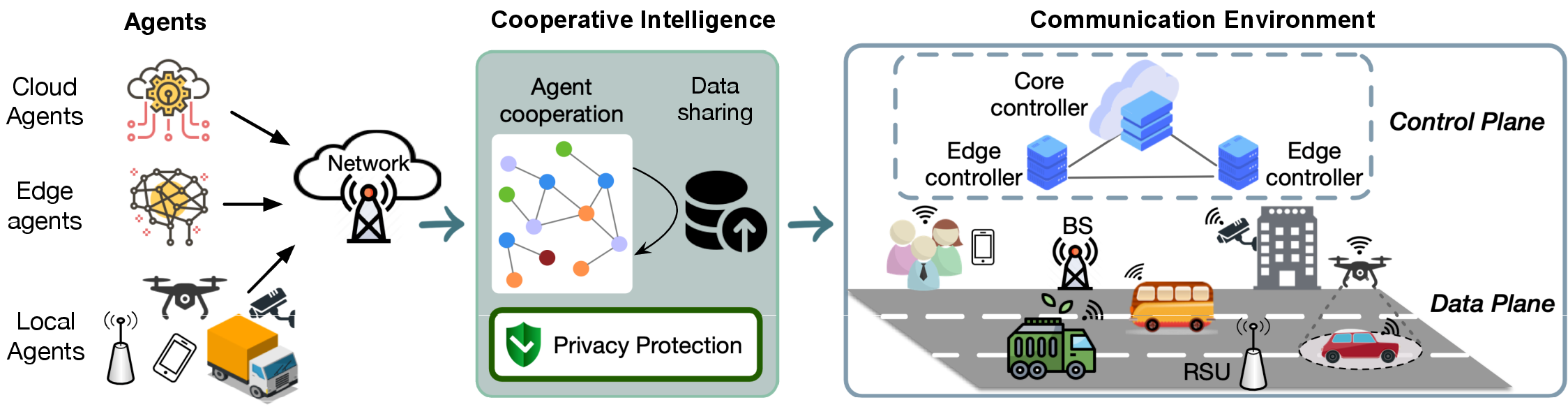}
\caption{Cooperative intelligence for communications with case studies.}
\label{fig:Envs}
\end{figure*}

Several approaches exploit MARL with (partial) privacy protection.
However, they either lack efficiency by introducing high overheads \cite{zhuo2019federated} or provide inadequate privacy protection (e.g., vulnerable to inference attacks \cite{peng2021facmac}).
Furthermore, implementing privacy protection techniques such as data encryption and federated learning to existing MARL (e.g., FedQ \cite{zhuo2019federated}) introduces the notable overheads (e.g., computation and bandwidth cost).
In this paper, to make MARL deployable in practical communication problems, we propose PP-MARL, a novel cooperative MARL scheme that 

\begin{enumerate}
\item improves privacy protection for CI by storing privacy-sensitive data (e.g., agents' observations and actions) locally and encoding raw data for privacy-preserving sharing;
\item prevents inference and differential attacks by exploiting homomorphic encryption (HE) \cite{alaya2020homomorphic} and differential privacy (DP) \cite{hassan2019differential} on the shared information;
\item reduces the communication and computation overheads by introducing hierarchical critics via split learning, which is designed as an HE-friendly architecture.

\end{enumerate}
Moreover, we evaluate PP-MARL in two representative use cases: mobility management in drone-assisted communication and network control with edge intelligence.
Simulation results reveal that PP-MARL can achieve better collaboration, lower overheads and better privacy protection than existing schemes.

\section{MARL for CI in Communications}
\label{CI-MARL}
We provide an overview of MARL, MARL for agents' coordination within CI, challenges of applying MARL in communications through case studies, and existing privacy-preserving schemes.

\vspace{-1em}
\blue{
\subsection{MARL for CI}
\label{ppMARL}

CI is an integral function in multi-agent and distributed systems, where multiple intelligent agents seek collaborations to improve their welfare jointly.
Many-fold entities can be regarded as intelligent agents in communication scenarios, including local, edge, and cloud agents (see Fig.~\ref{fig:Envs}).

One prominent way of coordinating agents to achieve CI in the long term is MARL \cite{zhang2019multi}.
MARL can be modeled as a Markov decision process (MDP).
\textbf{Agents} in MARL learn action making by interacting with the environment.
At each time, each agent observes the current state of the environment from its personal perspective (called as \textbf{observations}) and makes a proper \textbf{action} according to its knowledge.
Then, the environment returns a \textbf{reward} as feedback, and moves to the next state according to the transition probability.
In this process, the stored data of each agent is called experiences, including observations, actions, rewards, and successive observations.
Furthermore, cooperative MARL is designed to find cooperative policies of agents by jointly evaluating their performance.
Cooperative policies have advantages over independent strategies, especially when the agents can obtain only partial observations.

There are many algorithms that can train agents to approach their goals, which is to find an optimal policy that maximizes long-term and overall return expectations.
The most famous one is actor-critic.
The \textbf{actor} located in an agent is in charge of action-making given observations of environment.
The \textbf{critic} is a function that maps an observation-action pair to a scalar value (also called as \textbf{Q values}) representing the expected total long-term rewards that the agent is expected to accumulate.}{}

\begin{figure}[h]
\centering
\includegraphics[width=0.44\textwidth] {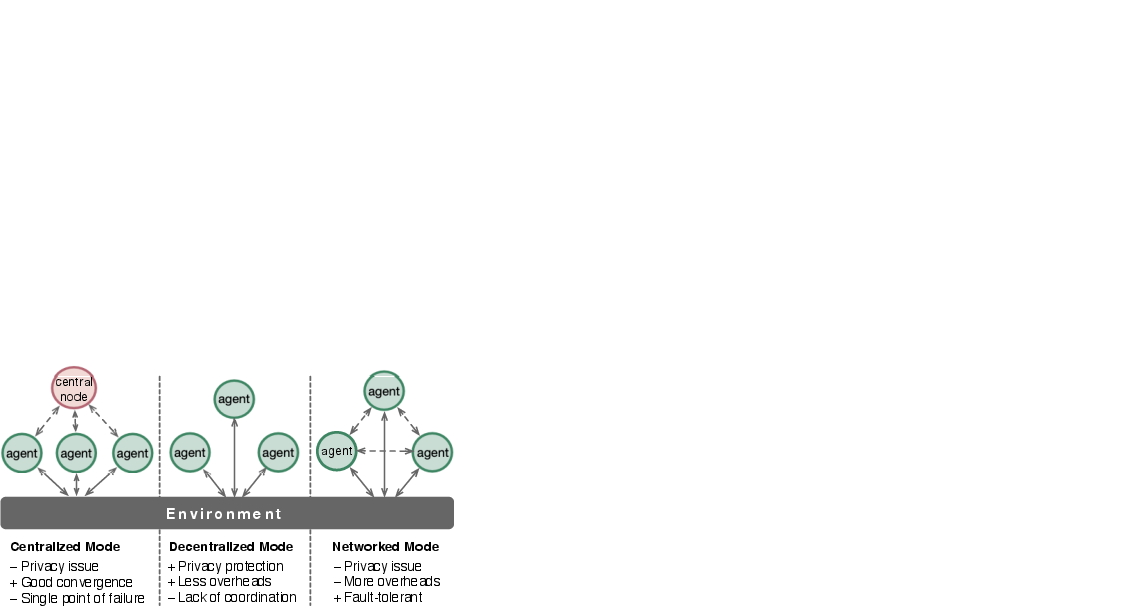}
\caption{Agent cooperation modes.}
\label{fig:arch}
\end{figure}

In MARL, three cooperation modes promote agent collaboration through information sharing and cooperative learning, as shown in Fig.~\ref{fig:arch}. Firstly, the \emph{centralized mode}, exemplified by MADDPG \cite{lowe2017multi} and FACMAC \cite{peng2021facmac}, relies on a central node for coordination. This mode, while effective, hinges on a trusted central authority, posing potential single-point failure and privacy risks due to data sharing.
Secondly, in the \emph{decentralized mode}, agents are independent learners and workers with no explicit information exchanged with each other, such as DDPG \cite{lillicrap2015continuous}.
Strictly speaking, this mode is non-collaborative and its partial observation environments may lead to non-convergence of the learning \cite{lowe2017multi}.
Lastly, the \emph{networked mode} is a decentralized model with networked agents, in which agents are connected via a communication network so that their local information can be spread across the network, such as FedQ \cite{zhuo2019federated}.
However, this mode scales poorly as the number of agents increases, because its communication overheads increase dramatically.
Besides, information sharing among agents poses privacy issues.
\blue{Hybrid models have also emerged in recent years, for example, decentralized execution by agents (i.e., make local actions based on local observations) with centralized training using the central node, such as MADDPG \cite{lowe2017multi} and FACMAC \cite{peng2021facmac}.}{}

\vspace{-0.5em}


\subsection{Use Case Studies}
\label{sec:usecases}
To understand the challenges of applying MARL-assisted CI in communication, we study two typical multi-agent use cases as illustrated in Fig.~\ref{fig:Envs}.
Notably, privacy issues arise in various CI-related scenarios, and we select two of the most popular topics for illustration.

\subsubsection{Mobility management in drone-assisted communication}
In this use case, drones, working as aerial communication infrastructures, provide network services to targets (e.g., vehicles and users).
Each drone has limited coverage and cannot serve all targets anytime and anywhere.
Therefore, multiple drones need to cooperate to cover targets while avoiding collisions.
Drones, also working as agents, can observe the positions of targets and other drones in range and then decide on movements to maximize the joint rewards (e.g., the number of covered targets) cooperatively.
Since each drone has a partial observation of the environment, it would be difficult to make a wise movement without cooperation.
However, information sharing to support cooperation brings a privacy risk (e.g., revealing the location of vehicles and the movement of drones) and introduces additional communication overheads (e.g., bandwidth and energy consumption).

\subsubsection{Network control with edge intelligence}
We study CI for network control (e.g., configure routing tables and bandwidth allocation) in hierarchical software-defined IoV (HSD-IoV) \cite{yuan2020dynamic}.
HSD-IoV consists of edge controllers located in base stations (BSs) and a core controller in the cloud (see Fig.~\ref{fig:Envs}).
The edge controllers are close to vehicles and offer low-latency responses, while the core controller coordinates edge controllers' learning.
Edge controllers need to adaptively and cooperatively assign the load and eventually provide overall low-latency responses.
The edge controllers, working as agents, can obtain requests from vehicles in their coverage range, and then cooperatively decide how to assign the management of vehicles to edges. 
The detailed formulations can be found in our previous work \cite{yuan2020dynamic}.
Data sharing for cooperation introduce privacy issues, given that the other controllers may not be trusted or belong to different companies.
For example, vehicles are susceptible to tracking and profiling due to their exposed locations and communication demands to the untrusted.
Besides the overheads analyzed in the first use case, agent cooperation during action-making (i.e., execution) results in delays which cannot be ignored, as network control is a delay-sensitive task.

\noindent
\textbf{Challenges.}
As analyzed above, CI has the potential to improve communications, but it also brings new challenges in privacy protection and overheads.
Given the practical constraints (e.g., privacy concerns and limited bandwidth), the CI used for communication problems needs to be efficient privacy protected, which includes
(1) privacy protection over agents' experiences (e.g., locations of vehicles in observations of use case 2);
(2) low overheads for cooperative learning and execution (e.g., bandwidth, energy, latency and computation);
and (3) ensuring the gains of collaboration.

\vspace{-1em}
\subsection{Existing Privacy-Preserving Schemes for MARL}
\label{State-of-the-art}

Three kinds of MARL schemes have recently been proposed to address the privacy challenges.
\subsubsection{Federated learning (FL)}
FL \cite{li2020federated} can protect the local data, allowing multi-agents to train with local data and build a shared model by sharing the model parameters and training gradients.
Notably, FL requires agents to be homogeneous, e.g., the neural networks have the same architecture and the same feature space.
Given agents in MARL is often heterogeneous (e.g., with different observation range), introducing FL to MARL will increase the size of the model and communication data.
For example, FedQ \cite{zhuo2019federated} federatively builds global models using networked cooperation modes.
FedQ incurs higher communication overheads and delays in agents' execution because its training and execution rely on information from neighbour agents.
Therefore, the performance of FedQ is not acceptable in time-sensitive scenarios.


\subsubsection{Data encoding}
It is the process of converting the raw data into a specific format for privacy protection.
Existing schemes use shallow neural networks to encode raw data, such as FedQ and FACMAC.
In FedQ, the encoded data are shared among agents to build global models.
FAMCMAC, \Repthree{a decomposition-based MARL,}{1}{} shares the encoded data to the central node to build centralized but factored critics under the assumption that the central node knows the state of the overall environment.
However, in practical problems, this assumption relies on the sharing of raw data (i.e., the agent's observations) and poses serious privacy concerns.
Therefore, existing schemes either bring high overheads (e.g., FedQ) or have serious privacy issues (e.g., FACMAC).

\begin{figure*}[h!]
\centering
\includegraphics[width=0.8\textwidth] {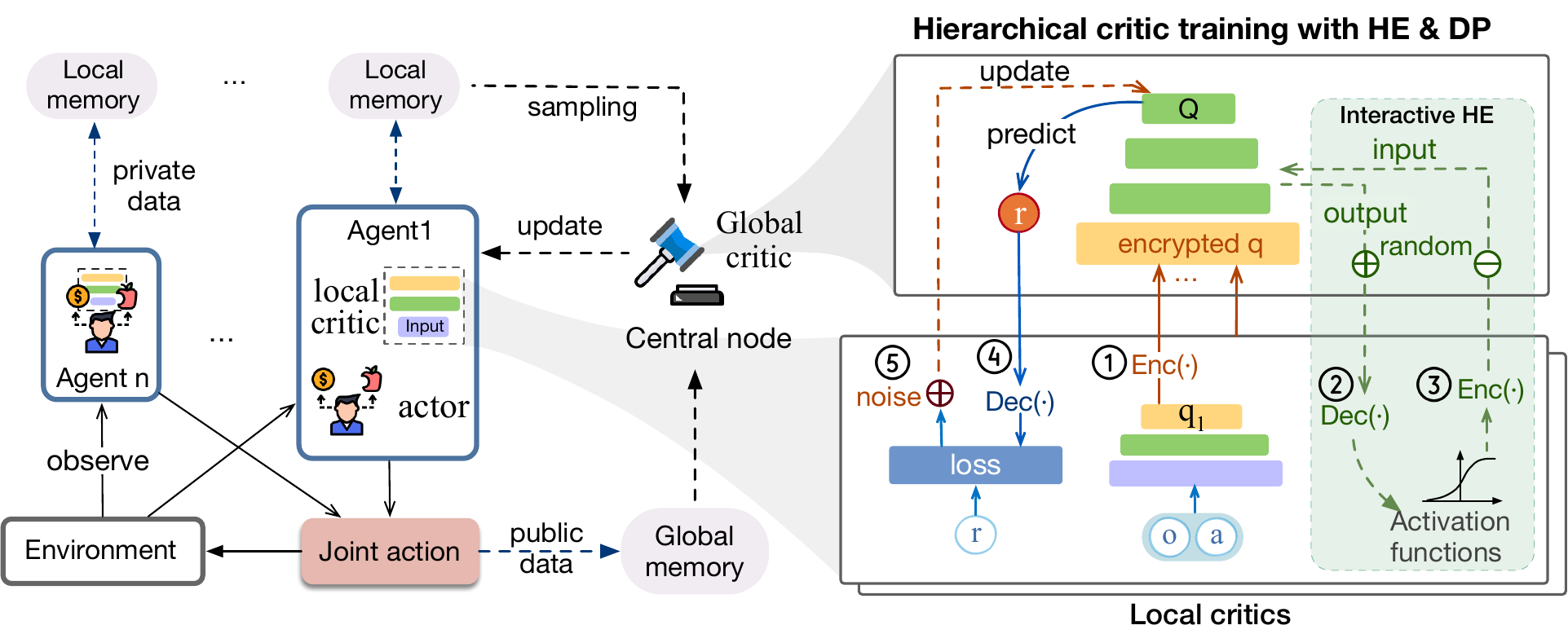}
\caption{PP-MARL: distributed execution and centralized learning with hierarchical critics and memories. \blue{HE is used to offer protection over q values; DP prevents the inference of r from the loss value.}{} o, a, r, q, Q stand for observations, actions, rewards, q values and Q values.}
\label{fig:Fed}
\end{figure*}

\noindent
\subsubsection{Data encryption}
This approach converts raw data to encrypted data, such as HE \cite{suh2021sarsa} and DP \cite{jiang2021differential}.
HE has been introduced for data encryption in single-agent reinforcement learning \cite{suh2021sarsa}.
HE excels in secure computations on encrypted data, preserving data privacy throughout. In contrast, traditional encryption, designed for data confidentiality during storage and transmission, requires decryption before computation, potentially exposing sensitive data. Moreover, compared to non-decryption methods like Secure Multi-Party Computation, HE is better suited for complex mathematical operations on encrypted data.
However, the efficiency issue is the most challenging in deploying HE for MARL because the HE scheme incurs significant computational overheads for supporting encryption on the shared data.
The overhead is incredibly high for traditional MARL due to their high volume of shared data (e.g., all experiences in MADDPG and observations in FACMAC).
Besides, DP \cite{jiang2021differential} is another technique that protects data by injecting noises.
DP allows statistical analysis of the data set (e.g., mode and mean) and does not reveal whether the individual's information was included in the original data set or not.

\section{Key Design of PP-MARL}
As mentioned above, the existing schemes cannot simultaneously address all the challenges mentioned in \S \ref{sec:usecases} as they either lack collaboration, have inadequate privacy protection, or introduce high overheads.
We propose PP-MARL to offer efficient privacy protection for CI given the practical challenges in communications.
In a nutshell, \blue{PP-MARL is designed to allow decentralized execution and centralized training with privacy protection via introducing \emph{hierarchical critics and memories}}{}, as presented in Fig.~\ref{fig:Fed}.
(1) \blue{Decentralized execution means that agent can make actions without the intervention of anyone else.}{}
It can offer low latency by the independent decision making of agents, which meets the requirements of latency-sensitive applications in communication; 
(2) centralized training with a central node for coordination training guarantees convergence;
(3) hierarchical critics and memories support efficient privacy protection by encoding raw data to a lower-dimension space, which can protect sensitive data and reduce overheads when introducing HE and DP to prevent inference and differential attacks.

\vspace{-1em}
\subsection{Execution and Training in PP-MARL}

The execution is an interactive process between agents and the environment, during which agents can get and store data (i.e., experience) for learning.
The execution is designed to be decentralized in PP-MARL, enabling low-latency actions by decoupling dependencies from other agents.
Firstly, agents independently (and possibly partially) observe the environment and obtain local observations (e.g., the position of vehicles within the observation range).
Next, each actor in the agents makes an action decision given its local observation.
All the actions are aggregated into joint actions, and then the joint actions act on the environment (e.g., drones move to target places and offer services to vehicles).
Then, the environment gives rewards to the agents given their actions and environment state, and the environment transfers to new states.
Lastly, agents' experiences are hierarchically stored for privacy protection.
The global memory stores public data (i.e., memory identifier); 
each agent stores private experiences into its local memory, including local observations, individual actions, rewards, and successive observations.

The centralized training process for hierarchical critics and actors is described below.
(1) Hierarchical critics are updated by minimizing the loss function over Q values.
Firstly, the central node provides sampled identifiers to all involved agents for memory extraction.
Secondly, local critics give corresponding q values to the global critic. Next, the estimated rewards of each agent are calculated by the central node, which is the distance between Q values and discounted target critic values (i.e., the estimated further accumulated rewards) of the next step. The estimated rewards are then sent back to agents. 
After receiving the estimated rewards, each agent calculates the loss defined as the mean squared error between the estimated rewards and the locally stored rewards (i.e., the ground truth) and sends the loss back to the central node.
Finally, the hierarchical critic is updated with backward propagation using the loss.
\Repthree{The local critics then update their parameters by receiving gradients from the global critic through backpropagation, which applies the chain rule.}{3}{}
(2) The parameters of the actor-networks are updated using the sampled policy gradient \cite{lowe2017multi}. 
Policy gradient methods optimize parameterized actors concerning the Q values by gradient descent.
The Q value is obtained by merging q values through global critic networks.


\vspace{-1em}
\subsection{Privacy Protection in PP-MARL}
\label{privacy_arc}

To avoid privacy leakage in PP-MARL, we propose a 2-stage privacy protection scheme.

\subsubsection{Raw data encoding}
\blue{To evaluate the overall performance of agents (i.e., to calculate the Q-values), the central node needs access to information of agents (e.g., experiences in MADDPG).
To avoid sharing of raw data, we introduce data protection by encoding the experiences.
}{}
Specifically, we propose hierarchical critics to estimate the value of joint action-values with privacy protection.
A hierarchical critic consists of local critics of the involved agents and a global critic for local value integration.
The local critic in each agent evaluates local performance (i.e., q values) on the local observations and actions.
By local critic, we transfer the experiences into q values, \blue{which are used to evaluate local actions made by actors.
Sharing q values to the central node for global evaluation can avoid raw data sharing and also ensure the accuracy of the overall evaluation.}{}
Strictly speaking, a local critic is a utility function instead of a value function because it cannot estimate expected rewards by itself, and therefore q values cannot be used for updating the actors directly.
The global critic is located in the central node and designed as a deep neural network.
It combines q values of all involved agents to evaluate the joint actions based on the joint observations and gives global critic values (abbreviated as Q values).
As agents may have different reward functions, multiple global critics can be introduced for different agents or teams in the central node.
With hierarchical critics, as explained previously, participated agents’ privacy is preserved through information embedding and excluding the need for direct and repeated exchange of private information between agents and the central node.

The hierarchical critic differs from a decentralized critic (e.g., DDPG) and a centralized critic (e.g., MADDPG).
The former estimates the local action-values based only on local observations and actions for each agent.
On the other hand, the latter can use the information of all agents to estimate the joint action-values based on the joint data (i.e., experiences).
However, the centralized critics introduce high communication overheads and privacy issues.
For example, the critics in MADDPG are trained with experiences that include observations, actions, and rewards of all agents.
To avoid high overheads and privacy leakage in training, PP-MARL introduces hierarchical critics so that only q values are shared to estimate the joint action-values.

\subsubsection{Preventing inference and differential attacks} 
\blue{
Even with data encoding, PP-MARL still faces privacy issues in two aspects.
Firstly, the q values may leakage the local observations and the actions.
Secondly, 
the loss is statistical results of a batch of data (i.e., the samples selected by the global critic) that faces differential attack.}{}
To avoid inferring private data through the q values and the loss, we introduce HE and DP to improve privacy protection further, as depicted in Fig.~\ref{fig:Fed} (the right part).

Deploying HE to neural networks meets challenges because HE schemes support only homomorphic arithmetic operations such as homomorphic addition and multiplication.
However, most of the popular and standard activation functions in neural networks are non-arithmetic, such as ReLU and sigmoid \cite{alaya2020homomorphic}.
To address this issue, we introduce interactive HE, which executes non-arithmetic operations (i.e., activation functions) locally by agents or a trusted device with the secret key.
The interactive HE operations for ensuring privacy during information exchange include: \ding{172} encrypting q values, \ding{173} decrypting the outputs of middle layers for activation, \ding{174} encrypting the data after activation and sending back to the global critic, \ding{175} decrypting predicted rewards to calculate losses, which will be used for updating the critics\blue{, and \ding{176} adding Gaussian noise (i.e., DP) over the loss to offer differential privacy protection for the rewards.}{}
\Reptwo{To avoid privacy leaks when sharing the outputs of middle layers with agents for activation, we randomly partition the outputs among multiple agents and insert randomly generated values to increase the complexity of inferring private data.
Once the global critic receives the returned values after activation, it can remove the added randomness to reconstruct the data.}{4}{}
Clearly, this series of processes introduces additional overheads (e.g., bandwidth and energy costs) in data encryption, decryption and transmission.
One approach to solving the above challenge is to replace the non-arithmetic activation function with the low order polynomials or linear activation function \cite{gilad2016cryptonets}.
With such a mitigation approach, the encrypted data can be processed by the global critic alone, and only the process \ding{172}, \ding{175}, and \ding{176} are needed.
Therefore, the overheads in PP-MARL will be much lower without the need to decrypt, encrypt, and transfer the outputs from hidden layers (i.e., the processes with green dash lines).
However, this approach brings a decline in cooperative learning.


To reduce the overheads, PP-MARL is designed to be HE-friendly by reducing the amount of shared data (i.e., q values in low dimension) that needs to be encrypted compared with existing schemes.
In contrast, if HE is used for protecting the shared data in MADDPG, it needs to encrypt over observations, actions, and rewards; FACMAC needs to encrypt over q values and states.

\section{Evaluation and Analysis}
\label{sec:sim}
We compare PP-MARL with existing approaches in the two use cases discussed in \S \ref{sec:usecases} assessing learning performance, privacy protection, and overheads of communication and computation.

\vspace{-1em}
\subsection{Drone-assisted Communication}
\label{sec:sima}
In this experiment, three drones are employed to cover three target places and provide communication services. The drones serve in a rectangular space of 1$\times$1 km$^2$. The positions of drones and targets are initialized randomly. Each target place is a circular space with a radius of 40 m, and the coverage radius of drones is 50 m.
\Repthree{The drones are set to fly at a fixed height of 50 meters and with a bandwidth of 1 MHz. For further details on the setup, please refer to \cite{yuan2021harnessing}.}{2}{}
Each episode is divided into 10 identical time intervals (i.e., steps).
Drones aim to cover all the targets as soon as possible cooperatively.


\begin{figure}[h!]
\centering
\includegraphics[width=0.39\textwidth] {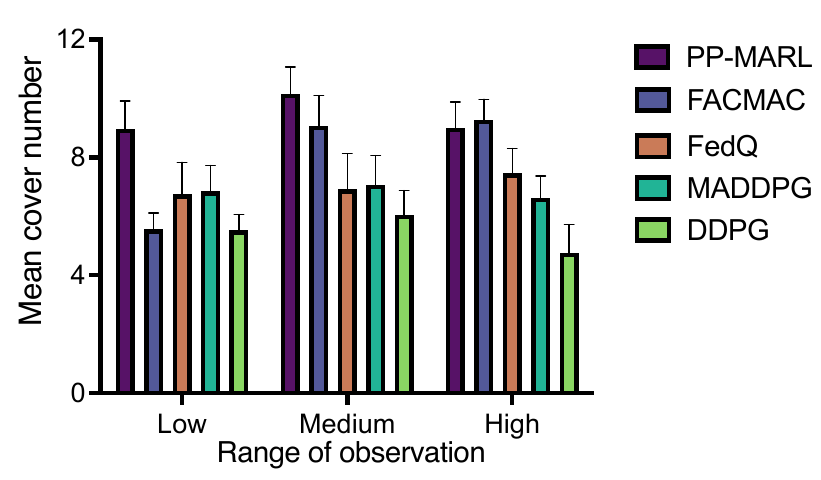}
\caption{Drone-assisted communication: mean number of target places covered by drones (per episode) in different observation ranges.}
\label{fig:rewards}
\end{figure}

Fig.~\ref{fig:rewards} shows the learning performance comparison with four representative algorithms.
The y-axis is the mean number of target places covered by drones in each episode, and the x-axis is the ability of agents' observations, which ranges from low to high.
We find that cooperative learning schemes (i.e., PP-MARL, FACMAC, FedQ, and MADDPG) generally achieve better performance than non-cooperative schemes (i.e., DDPG).
While individual intelligence alone may be limited, collaboration in learning can bring together agents' intelligence, leading to better performance for the application.
Generally, PP-MARL achieves better performance than the others.
\Repone{PP-MARL performed slightly worse than FACMAC in scenarios with a high observation range, likely due to PP-MARL being more susceptible to redundant observations (i.e., leading to longer training time).}{2}{}
Furthermore, by comparing the different observation ranges, we can find that a broader range of observations does not necessarily lead to better performance for all MARL algorithms.
A wider range of observations can provide more information for the agents in acting while bringing difficulty in training owing to the larger input dimension.

\begin{table*}[h!]
  \caption{Index comparison: cooperation mode, private data, public data, privacy protection, learning performance (named as gains), and overheads, including bandwidth (in MB), energy, and computation relative to DDPG.
  w stands for weights of neural networks.
  Data with a tilde is encrypted.
  h is the coefficient for HE operation cost relative to the multiplication operation.
}
  \label{table:Diversity}
  \centering
    \resizebox{18.2cm}{!} 
{ 
  \begin{tabular}{c c  c c c c c r l c c}
    \toprule
   \textbf{Approach} & \textbf{Mode} & \textbf{Private} & \textbf{Public} & \textbf{Protection}  & \textbf{Gains} & \textbf{Bandwidth} & \multicolumn{2}{c}{\textbf{Energy}} & \textbf{Computation}  & \textbf{Delay}\\ 
    \midrule
     DDPG & Decentral  & o, a, r & $\varnothing$ & \textbf{0.276}, 0.044  &5.5 & \textbf{0}  &  \textbf{1}  , & \textbf{1} &  \textbf{1} & \textbf{1}\\ 
     MADDPG & Central & $\varnothing$ & o, a, r & 0, 0 & 6.9 & 0.34 & 1x,& 14x & 1.33x & 1x\\ 
     FedQ & Networked & o, a & r, q, w & 0.185, 0.041  & 7.0 & 0.32 & 8.3x,& 13.3x &  1.63x &2.2x\\ 
     FACMAC & Central & a  & o, r, q & 0.099, 0.015  & 8.0 & 0.19 & 1x,& 9.28x & 1.48x &1x\\ 
    \textbf{PP-MARL w/o HE}& Central & o, a, r & q & \underline{0.191}, 0.041  & \textbf{9.4}& \underline{0.03} & 1x,& \underline{1.96x} & \underline{1.01x} &1x\\ 
    \midrule
     MADDPG-\textbf{HE$^{\text{1}}$} & Central & $\varnothing$ & \~{o}, \~{a}, \~{r} &  0.276, 0.052 & 6.9 & 3.3 & 1x,& 128x & (1.3+0.55h)x &1x\\
    \textbf{PP-MARL-HE$^{\text{1}}$}& Central & o, a, r & \~{q} & 0.276, 0.052  & 9.4& 1.03 & 1x,& 40x & (1.0+0.16h)x &1x\\ 
    \textbf{PP-MARL-HE$^{\text{2}}$}& Central & o, a, r & \~{q} & 0.276, 0.044  & 8.8 & 0.03 & 1x,& 2x & (1.0+0.02h)x &1x\\ 
    \bottomrule
  \end{tabular}
  }
  \vspace{-0.8em}
\end{table*}

\vspace{-1em}
\subsection{Network Control with Edge Intelligence}
\label{sec:simb}
For HSD-IoV, we use the dataset of vehicle mobility traces from SNDlib (data can be found in \cite{yuan2020dynamic}).
This dataset offers real-time position data reported by buses from Rio de Janeiro (7 days in $24$-hour format).
The region is a 10$\times$10 km$^2$ map of Rio de Janeiro, Brazil.
There are 4 BSs equipped with edge controllers, with a coverage radius of 4 km.
The edge controllers' observations vary based on their locations, and their service coverage may overlap.
We show simulation results of PP-MARL compared with FACMAC, MADDPG, DDPG, and two non-intelligent approaches (i.e., DB and RC \cite{yuan2020dynamic}).
DB is a distance-based greedy policy, and RC is centralized intelligence using a cloud controller. 
As value-based MARL algorithms, such as FedQ, cannot be used for problems with continuous action space, we cannot evaluate their performance for this problem.

\begin{figure}[h!]
\centering
\includegraphics[width=0.4\textwidth] {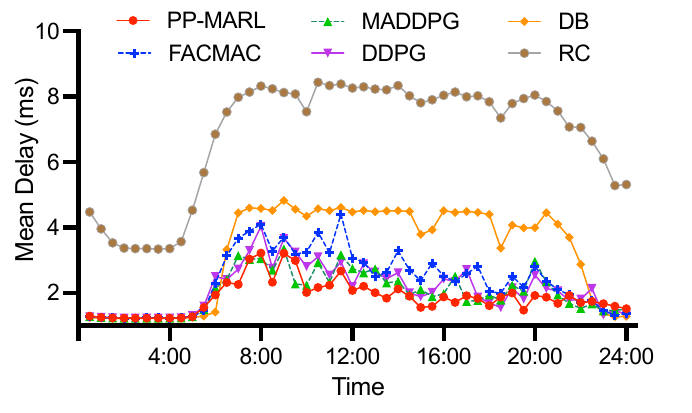}
\vspace{-0.5em}
\caption{Network control with edge intelligence: mean delay over time. PP-MARL can reduce delay than the others, which is more obvious during rush hours.}
\label{fig:delay}
\end{figure}

Fig.~\ref{fig:delay} presents the mean control delays of vehicles varied in 24 hours.
Among all schemes, PP-MARL has the shortest delay compared to the others, i.e., 8.5\%, 12.1\%, 21.5\%, 43\%, and 72.6\% shorter than MADDPG, DDPG, FACMAC, DB, and RC on average, respectively.
By comparing RC with other methods, it can be demonstrated that edge intelligence has less control delay than centralized intelligence (i.e., RC).
DB has a higher delay than the MARL algorithms, as it only considers distance and ignores the load of edge controllers.
DDPG has limited delay reduction compared with cooperative MARL algorithms due to a lack of cooperation.
In summary, CI can provide flexible network management, and PP-MARL can further reduce control delays through efficient cooperation.

\vspace{-1em}
\subsection{Privacy Protection and Overheads}
We compare the privacy protection and overheads of PP-MARL with baselines for use case 1 in Table \ref{table:Diversity}.
Non-intelligent methods, such as DB and RC, are not big data-driven methods and do not introduce obvious overheads and privacy issues, and we do not show their comparison here.
\blue{The noises added on the loss follow the Laplace distribution with center 0 and scale 0.01, which can preserve (0.1, 0)-DP given rewards are normalized.}{}

Firstly, the \emph{privacy protection degree} in Table \ref{table:Diversity} is defined as the root mean square error (RMSE) between the inferred data and actual ones.
\blue{For example, the locations of targets and the movement of drones are the private data of use case 1, and the RMSE is the difference between the inferred locations and movement and actual ones.}{}
A higher RMSE implies that it is more difficult to infer private data from public data and thus able to provide better privacy protection.
The results reveal that DDPG provides the best protection because it is non-cooperative without information sharing, and the inferred data are predicted from zero vectors as inputs.
However, as our experimental results in \S \ref{sec:sima} and \S \ref{sec:simb} show, non-cooperative MARL (i.e., DDPG) can be less effective with lower gains.
PP-MARL provides better privacy protection than the other cooperative methods because it only shares q values.
Although value-decomposition methods (e.g., FACMAC) can protect privacy for actions, private data may still be inferred much easier from public data.
FedQ provides marginally less privacy protection than PP-MARL because it shares rewards in addition to q values.
\Reptwo{Finally, we analyze privacy protection for inferring other agents' private data from each agent, as shown in the second column values of protection in Table \ref{table:Diversity}.
These values are significantly lower than the first-column values because overlapping observations among agents make it easier to infer from agents rather than the central node.
Next, we can see that HE performs better than the non-HE cooperative methods.
In our interactive HE (HE$^{\text{1}}$), sharing the outputs of middle layers does not introduce additional privacy leakage, and the use of randomness provides better protection against inference attacks compared to HE$^{\text{2}}$, which is a non-interactive HE that uses linear activation.}{3}{}


Secondly, we compare PP-MARL with baselines in terms of bandwidth \blue{(the size of messages to be delivered, in MB), energy consumption}{}, computation, and delays.
\blue{The energy consumption includes the computational operation of the neural network (mainly multiplication and addition operations) and the energy consumed by the transmission.}{}
The results show that PP-MARL has obviously low overheads than the other cooperative MARL.
The results are consistent with our previous analysis in \S \ref{State-of-the-art}: DDPG has no additional overheads as it is non-cooperative; MADDPG introduces higher bandwidth due to more data sharing between agents and the central node; FedQ has a disadvantage in supporting latency-sensitive applications because of higher delays and energy cost in execution, and clearly, it introduces high communication and computation overheads; FACMAC has lower overheads than FedQ and MADDPG, but is still significantly higher than PP-MARL due to the shared states.

Thirdly, we show the performance and overheads of HE and DP deployed in PP-MARL and MADDPG.
The results show that introducing interactive HE into MARL brings obviously high overheads during training.
For example, MADDPG with HE$^{\text{1}}$ introduces around 10 times higher bandwidth cost and 9 times more energy consumption during training.
In PP-MARL, the increment by introducing interactive HE cannot be ignored, even though PP-MARL is designed to be HE-friendly by reducing native communication data volume (only sharing q values) and decomposing the critic neural network to be hierarchical.
While PP-MARL-HE$^{\text{2}}$ has a lower overhead compared to PP-MARL-HE$^{\text{1}}$, its cooperative performance is reduced with fewer gains.
Moreover, we prefer interactive HE only when the agents (e.g., edge and end devices) are powerful.

From the comparison, we can draw the following conclusions.
PP-MARL provides the best privacy protection with the lowest overheads among all the cooperative MARL schemes.
Besides, we introduce HE and DP to further improve privacy protection at the cost of higher overheads in communication and computation.
In addition, PP-MARL proves to be HE-friendly, introducing less overhead than the other MARL schemes with HE.

\section{Conclusion and Discussion}
We exploited CI in facilitating next-generation networks, enabling agent collaboration through data sharing via communication.
However, privacy concerns and communication constraints hindered MARL deployment. To tackle this, we introduced PP-MARL, a privacy-preserving MARL scheme with an HE-friendly architecture. 
Simulation results revealed that cooperative learning promotes collaboration among agents, and PP-MARL can provide better privacy protection with lower overhead than state-of-the-art approaches.



\section*{Acknowledgment}
This work has been partly funded by Horizon Europe COVER project (No. 101086228) and the Alexander von Humboldt Foundation.

\bibliographystyle{IEEEtran}

\bibliography{ref}

\begin{IEEEbiographynophoto}{Tingting Yuan}
received her Ph.D. degree from Beijing University of Posts and Telecommunications (BUPT), China, in 2018. She held postdoctoral positions at INRIA, Sophia Antipolis, France, and at the University of Göttingen from 2018 to 2023. From 2023, she is working as Junior Professor at University of G\"ottingen.
\end{IEEEbiographynophoto}

\begin{IEEEbiographynophoto}{Hwei-Ming Chung} received his Ph.D. degree in Networks and Distributed Systems from University of Oslo, Norway in 2022. He is currently a cofounder of NooT (Zhuhai) Co., Ltd. His research interests include power system monitoring, smart grid, AI hardware acceleration and statistical signal processing.
\end{IEEEbiographynophoto}

\begin{IEEEbiographynophoto}{Xiaoming Fu (M'02, SM'09, F'22)} received the Ph.D. degree in computer science from Tsinghua University, China, in 2000. He is a Professor and the Head of the Computer Networks Group, University of G\"ottingen. He has also held visiting positions at ETSI, University of Cambridge, Columbia University, Tsinghua University, and UCLA. He is a Fellow of IEEE, a distinguished member of ACM, a fellow of IET, and a member of Academia Europaea.
\end{IEEEbiographynophoto}

\vspace*{\fill}

\end{document}